\documentclass[aps,prl,10pt,twocolumn,showpacs,superscriptaddress,floatfix,nobibnotes]{revtex4-1}

\usepackage{amsmath}
\usepackage{xspace}
\usepackage{graphicx}

\usepackage[tracking=true,kerning=true,expansion=true,spacing=false,factor=1100,stretch=20,shrink=20]{microtype}
\SetTracking{encoding={*}, shape=sc}{50} 

\usepackage{siunitx}
\DeclareSIUnit\clight{\text{\ensuremath{c}}} 

\usepackage[pdftex,unicode=true,pdfpagelabels=true,pdfusetitle,pdfauthor={The CDF Collaboration},hidelinks]{hyperref}

\setkeys{Gin}{width=\linewidth,keepaspectratio}

\newcommand{\figref}[1]{Fig. \ref{#1}\xspace}
\newcommand{\tabref}[1]{Table \ref{#1}\xspace}
\newcommand{\ttbar}{\ensuremath{t\bar{t}}\xspace}
\newcommand{\afb}{\ensuremath{A_\text{FB}}\xspace}
\newcommand{\ctt}{\ensuremath{\cos \theta_t}\xspace}
\newcommand{\dsigctt}{\ensuremath{{d\sigma}/{d(\ctt)}}\xspace}
\newcommand{\pythia}{{\sc pythia}\xspace}
\newcommand{\ppbar}{\ensuremath{p\bar{p}}\xspace}
\newcommand{\MET}{\ensuremath{\raisebox{.3ex}{\(\not\!\)}E_T}\xspace}
\newcommand{\METVEC}{\ensuremath{\raisebox{.3ex}{\(\not\!\)}{\vec E}_T}\xspace}

\begin{document}

\title{\texorpdfstring{\boldmath{Measurement of the Differential Cross Section
			\dsigctt for Top-Quark Pair Production in \ppbar Collisions at \(\sqrt{s} =
\SI{1.96}{\tera\eV}\)}}{Measurement of the Differential Cross Section
dsigma/d(cos theta\_t) for Top-Quark Pair Production in p-pbar Collisions at
sqrt(s) = 1.96 TeV}}

\affiliation{Institute of Physics, Academia Sinica, Taipei, Taiwan 11529, Republic of China}
\affiliation{Argonne National Laboratory, Argonne, Illinois 60439, USA}
\affiliation{University of Athens, 157 71 Athens, Greece}
\affiliation{Institut de Fisica d'Altes Energies, ICREA, Universitat Autonoma de Barcelona, E-08193, Bellaterra (Barcelona), Spain}
\affiliation{Baylor University, Waco, Texas 76798, USA}
\affiliation{Istituto Nazionale di Fisica Nucleare Bologna, \ensuremath{^{ii}}University of Bologna, I-40127 Bologna, Italy}
\affiliation{University of California, Davis, Davis, California 95616, USA}
\affiliation{University of California, Los Angeles, Los Angeles, California 90024, USA}
\affiliation{Instituto de Fisica de Cantabria, CSIC-University of Cantabria, 39005 Santander, Spain}
\affiliation{Carnegie Mellon University, Pittsburgh, Pennsylvania 15213, USA}
\affiliation{Enrico Fermi Institute, University of Chicago, Chicago, Illinois 60637, USA}
\affiliation{Comenius University, 842 48 Bratislava, Slovakia; Institute of Experimental Physics, 040 01 Kosice, Slovakia}
\affiliation{Joint Institute for Nuclear Research, RU-141980 Dubna, Russia}
\affiliation{Duke University, Durham, North Carolina 27708, USA}
\affiliation{Fermi National Accelerator Laboratory, Batavia, Illinois 60510, USA}
\affiliation{University of Florida, Gainesville, Florida 32611, USA}
\affiliation{Laboratori Nazionali di Frascati, Istituto Nazionale di Fisica Nucleare, I-00044 Frascati, Italy}
\affiliation{University of Geneva, CH-1211 Geneva 4, Switzerland}
\affiliation{Glasgow University, Glasgow G12 8QQ, United Kingdom}
\affiliation{Harvard University, Cambridge, Massachusetts 02138, USA}
\affiliation{Division of High Energy Physics, Department of Physics, University of Helsinki, FIN-00014, Helsinki, Finland; Helsinki Institute of Physics, FIN-00014, Helsinki, Finland}
\affiliation{University of Illinois, Urbana, Illinois 61801, USA}
\affiliation{The Johns Hopkins University, Baltimore, Maryland 21218, USA}
\affiliation{Institut f\"{u}r Experimentelle Kernphysik, Karlsruhe Institute of Technology, D-76131 Karlsruhe, Germany}
\affiliation{Center for High Energy Physics: Kyungpook National University, Daegu 702-701, Korea; Seoul National University, Seoul 151-742, Korea; Sungkyunkwan University, Suwon 440-746, Korea; Korea Institute of Science and Technology Information, Daejeon 305-806, Korea; Chonnam National University, Gwangju 500-757, Korea; Chonbuk National University, Jeonju 561-756, Korea; Ewha Womans University, Seoul, 120-750, Korea}
\affiliation{Ernest Orlando Lawrence Berkeley National Laboratory, Berkeley, California 94720, USA}
\affiliation{University of Liverpool, Liverpool L69 7ZE, United Kingdom}
\affiliation{University College London, London WC1E 6BT, United Kingdom}
\affiliation{Centro de Investigaciones Energeticas Medioambientales y Tecnologicas, E-28040 Madrid, Spain}
\affiliation{Massachusetts Institute of Technology, Cambridge, Massachusetts 02139, USA}
\affiliation{University of Michigan, Ann Arbor, Michigan 48109, USA}
\affiliation{Michigan State University, East Lansing, Michigan 48824, USA}
\affiliation{Institution for Theoretical and Experimental Physics, ITEP, Moscow 117259, Russia}
\affiliation{University of New Mexico, Albuquerque, New Mexico 87131, USA}
\affiliation{The Ohio State University, Columbus, Ohio 43210, USA}
\affiliation{Okayama University, Okayama 700-8530, Japan}
\affiliation{Osaka City University, Osaka 558-8585, Japan}
\affiliation{University of Oxford, Oxford OX1 3RH, United Kingdom}
\affiliation{Istituto Nazionale di Fisica Nucleare, Sezione di Padova, \ensuremath{^{jj}}University of Padova, I-35131 Padova, Italy}
\affiliation{University of Pennsylvania, Philadelphia, Pennsylvania 19104, USA}
\affiliation{Istituto Nazionale di Fisica Nucleare Pisa, \ensuremath{^{kk}}University of Pisa, \ensuremath{^{ll}}University of Siena, \ensuremath{^{mm}}Scuola Normale Superiore, I-56127 Pisa, Italy, \ensuremath{^{nn}}INFN Pavia, I-27100 Pavia, Italy, \ensuremath{^{oo}}University of Pavia, I-27100 Pavia, Italy}
\affiliation{University of Pittsburgh, Pittsburgh, Pennsylvania 15260, USA}
\affiliation{Purdue University, West Lafayette, Indiana 47907, USA}
\affiliation{University of Rochester, Rochester, New York 14627, USA}
\affiliation{The Rockefeller University, New York, New York 10065, USA}
\affiliation{Istituto Nazionale di Fisica Nucleare, Sezione di Roma 1, \ensuremath{^{pp}}Sapienza Universit\`{a} di Roma, I-00185 Roma, Italy}
\affiliation{Mitchell Institute for Fundamental Physics and Astronomy, Texas A\&M University, College Station, Texas 77843, USA}
\affiliation{Istituto Nazionale di Fisica Nucleare Trieste, \ensuremath{^{qq}}Gruppo Collegato di Udine, \ensuremath{^{rr}}University of Udine, I-33100 Udine, Italy, \ensuremath{^{ss}}University of Trieste, I-34127 Trieste, Italy}
\affiliation{University of Tsukuba, Tsukuba, Ibaraki 305, Japan}
\affiliation{Tufts University, Medford, Massachusetts 02155, USA}
\affiliation{University of Virginia, Charlottesville, Virginia 22906, USA}
\affiliation{Waseda University, Tokyo 169, Japan}
\affiliation{Wayne State University, Detroit, Michigan 48201, USA}
\affiliation{University of Wisconsin, Madison, Wisconsin 53706, USA}
\affiliation{Yale University, New Haven, Connecticut 06520, USA}

\author{T.~Aaltonen}
\affiliation{Division of High Energy Physics, Department of Physics, University of Helsinki, FIN-00014, Helsinki, Finland; Helsinki Institute of Physics, FIN-00014, Helsinki, Finland}
\author{S.~Amerio\ensuremath{^{jj}}}
\affiliation{Istituto Nazionale di Fisica Nucleare, Sezione di Padova, \ensuremath{^{jj}}University of Padova, I-35131 Padova, Italy}
\author{D.~Amidei}
\affiliation{University of Michigan, Ann Arbor, Michigan 48109, USA}
\author{A.~Anastassov\ensuremath{^{v}}}
\affiliation{Fermi National Accelerator Laboratory, Batavia, Illinois 60510, USA}
\author{A.~Annovi}
\affiliation{Laboratori Nazionali di Frascati, Istituto Nazionale di Fisica Nucleare, I-00044 Frascati, Italy}
\author{J.~Antos}
\affiliation{Comenius University, 842 48 Bratislava, Slovakia; Institute of Experimental Physics, 040 01 Kosice, Slovakia}
\author{G.~Apollinari}
\affiliation{Fermi National Accelerator Laboratory, Batavia, Illinois 60510, USA}
\author{J.A.~Appel}
\affiliation{Fermi National Accelerator Laboratory, Batavia, Illinois 60510, USA}
\author{T.~Arisawa}
\affiliation{Waseda University, Tokyo 169, Japan}
\author{A.~Artikov}
\affiliation{Joint Institute for Nuclear Research, RU-141980 Dubna, Russia}
% ...

\collaboration{CDF Collaboration}
\altaffiliation[With visitors from]{
\ensuremath{^{a}}University of British Columbia, Vancouver, BC V6T 1Z1, Canada,
\ensuremath{^{b}}Istituto Nazionale di Fisica Nucleare, Sezione di Cagliari, 09042 Monserrato (Cagliari), Italy,
\ensuremath{^{c}}University of California Irvine, Irvine, CA 92697, USA,
\ensuremath{^{d}}Institute of Physics, Academy of Sciences of the Czech Republic, 182~21, Czech Republic,
\ensuremath{^{e}}CERN, CH-1211 Geneva, Switzerland,
\ensuremath{^{f}}Cornell University, Ithaca, NY 14853, USA,
\ensuremath{^{g}}University of Cyprus, Nicosia CY-1678, Cyprus,
\ensuremath{^{h}}Office of Science, U.S. Department of Energy, Washington, DC 20585, USA,
\ensuremath{^{i}}University College Dublin, Dublin 4, Ireland,
\ensuremath{^{j}}ETH, 8092 Z\"{u}rich, Switzerland,
\ensuremath{^{k}}University of Fukui, Fukui City, Fukui Prefecture, Japan 910-0017,
\ensuremath{^{l}}Universidad Iberoamericana, Lomas de Santa Fe, M\'{e}xico, C.P. 01219, Distrito Federal,
\ensuremath{^{m}}University of Iowa, Iowa City, IA 52242, USA,
\ensuremath{^{n}}Kinki University, Higashi-Osaka City, Japan 577-8502,
\ensuremath{^{o}}Kansas State University, Manhattan, KS 66506, USA,
\ensuremath{^{p}}Brookhaven National Laboratory, Upton, NY 11973, USA,
\ensuremath{^{q}}Queen Mary, University of London, London, E1 4NS, United Kingdom,
\ensuremath{^{r}}University of Melbourne, Victoria 3010, Australia,
\ensuremath{^{s}}Muons, Inc., Batavia, IL 60510, USA,
\ensuremath{^{t}}Nagasaki Institute of Applied Science, Nagasaki 851-0193, Japan,
\ensuremath{^{u}}National Research Nuclear University, Moscow 115409, Russia,
\ensuremath{^{v}}Northwestern University, Evanston, IL 60208, USA,
\ensuremath{^{w}}University of Notre Dame, Notre Dame, IN 46556, USA,
\ensuremath{^{x}}Universidad de Oviedo, E-33007 Oviedo, Spain,
\ensuremath{^{y}}CNRS-IN2P3, Paris, F-75205 France,
\ensuremath{^{z}}Universidad Tecnica Federico Santa Maria, 110v Valparaiso, Chile,
\ensuremath{^{aa}}The University of Jordan, Amman 11942, Jordan,
\ensuremath{^{bb}}Universite catholique de Louvain, 1348 Louvain-La-Neuve, Belgium,
\ensuremath{^{cc}}University of Z\"{u}rich, 8006 Z\"{u}rich, Switzerland,
\ensuremath{^{dd}}Massachusetts General Hospital, Boston, MA 02114 USA,
\ensuremath{^{ee}}Harvard Medical School, Boston, MA 02114 USA,
\ensuremath{^{ff}}Hampton University, Hampton, VA 23668, USA,
\ensuremath{^{gg}}Los Alamos National Laboratory, Los Alamos, NM 87544, USA,
\ensuremath{^{hh}}Universit\`{a} degli Studi di Napoli Federico I, I-80138 Napoli, Italy
}
\noaffiliation

\date{\today}

\begin{abstract}

We report a measurement of the differential cross section, \dsigctt, for
top-quark-pair production as a function of the top-quark production angle in
proton-antiproton collisions at \(\sqrt{s} = \SI{1.96}{\tera\eV}\).  This
measurement is performed using data collected with the CDF~II detector at the
Tevatron, corresponding to an integrated luminosity of
\SI{9.4}{\per\femto\barn}.  We employ the Legendre polynomials to characterize
the shape of the differential cross section at the parton level.  The observed
Legendre coefficients are in good agreement with the prediction of the
next-to-leading-order standard-model calculation, with the exception of an
excess linear-term coefficient, \(a_1 = \num{0.40(12)}\), compared to the
standard-model prediction of \(a_1 = 0.15^{+0.07}_{-0.03}\).

\end{abstract}


\maketitle

The Collider Detector at Fermilab (CDF) and D0 experiments have measured an
anomalously large forward-backward asymmetry (\afb) in top-quark-pair (\ttbar)
hadroproduction.  The latest measurements are \(\afb =
\SI{16.4(45)}{\percent}\) from CDF~\cite{CDFLJAfb}, and \(\afb =
\SI{19.6(65)}{\percent}\) from D0~\cite{D0LJAfb}.  This asymmetry is the
manifestation of a charge asymmetry in \ttbar production via the
\textit{CP}-even~\footnote{\textit{CP} is the simultaneous transformation of
charge conjugation and parity.} initial state at the Fermilab Tevatron
proton-antiproton (\ppbar) collider.  The standard model
(SM)~\cite{Glashow:1961tr,*Weinberg:1967tq,*Salam:1968} predicts a small
forward-backward asymmetry, \SI{8.8(6)}{\percent}, at next-to-leading order
(NLO) in the strong coupling constant,
\(\alpha_s\)~\cite{BernSi,*BernPriv,KuhnRodrigo98}.  The tension between the
Tevatron measurements and the predictions has stimulated new work on the SM
calculation~\cite{Brodsky,HollikEWK,KuhnEWK,ManoharEWK} and on possible non-SM
sources for the asymmetry~\cite{BSMReviews}.  The charge asymmetry is also
under study at the LHC, but any effects are expected to be much smaller due to
the forward-backward symmetric (proton-proton) initial state~\cite{Berger}, and
the results are so far inconclusive~\cite{CMSAfb,ATLASAfb}.

We measure the differential cross section, \dsigctt, where \(\theta_t\) is the
angle between the top-quark momentum and the incoming proton momentum as
measured in the \ttbar center-of-mass frame.  The inclusive measurements of
\afb are equivalent to a two-bin measurement of this differential cross
section, with one bin forward (\(\ctt > 0\)) and one bin backward (\(\ctt <
0\)).  The full shape of the differential cross section provides additional
information, and has the potential to discriminate among various calculations
of the SM as well as models of non-SM physics.  One of the aims of this study
is to identify what aspects of the shape of \dsigctt explain the \afb.

We characterize the shape of \dsigctt by employing the Legendre
polynomials~\footnote{We normalize the Legendre polynomials so that
\(\frac{2\ell+1}{2} \int_{-1}^{1} d x P_{\ell}(x) P_{\ell'}(x)  =
\delta_{\ell \ell'}\), where \(\delta_{\ell \ell'}\) is the Kronecker delta.},
which are fundamental to the general theory of scattering of particles in the
present spin-averaged case~\cite{jacobwick}.  The orthonormality of these
polynomials on the interval \(\left[-1,1\right]\) allows a unique decomposition
of the cross section into a Legendre polynomial series.  We write
\begin{equation}\frac{d\sigma}{d(\ctt)} = \sum_{\ell=0}^{\infty} a_\ell
P_\ell(\ctt),\end{equation}where \(P_\ell\) is the Legendre polynomial of
degree \(\ell\), and \(a_\ell\) is the Legendre moment of degree \(\ell\).
Because the experimental sensitivity degrades as \(\ell\) increases, we
restrict the sum to \(\ell \leq 8\).  Since the moment \(a_0\) contains only
the total cross section, we scale all the moments so that \(a_0 = 1\).

\begin{figure}[htpb]
	\includegraphics{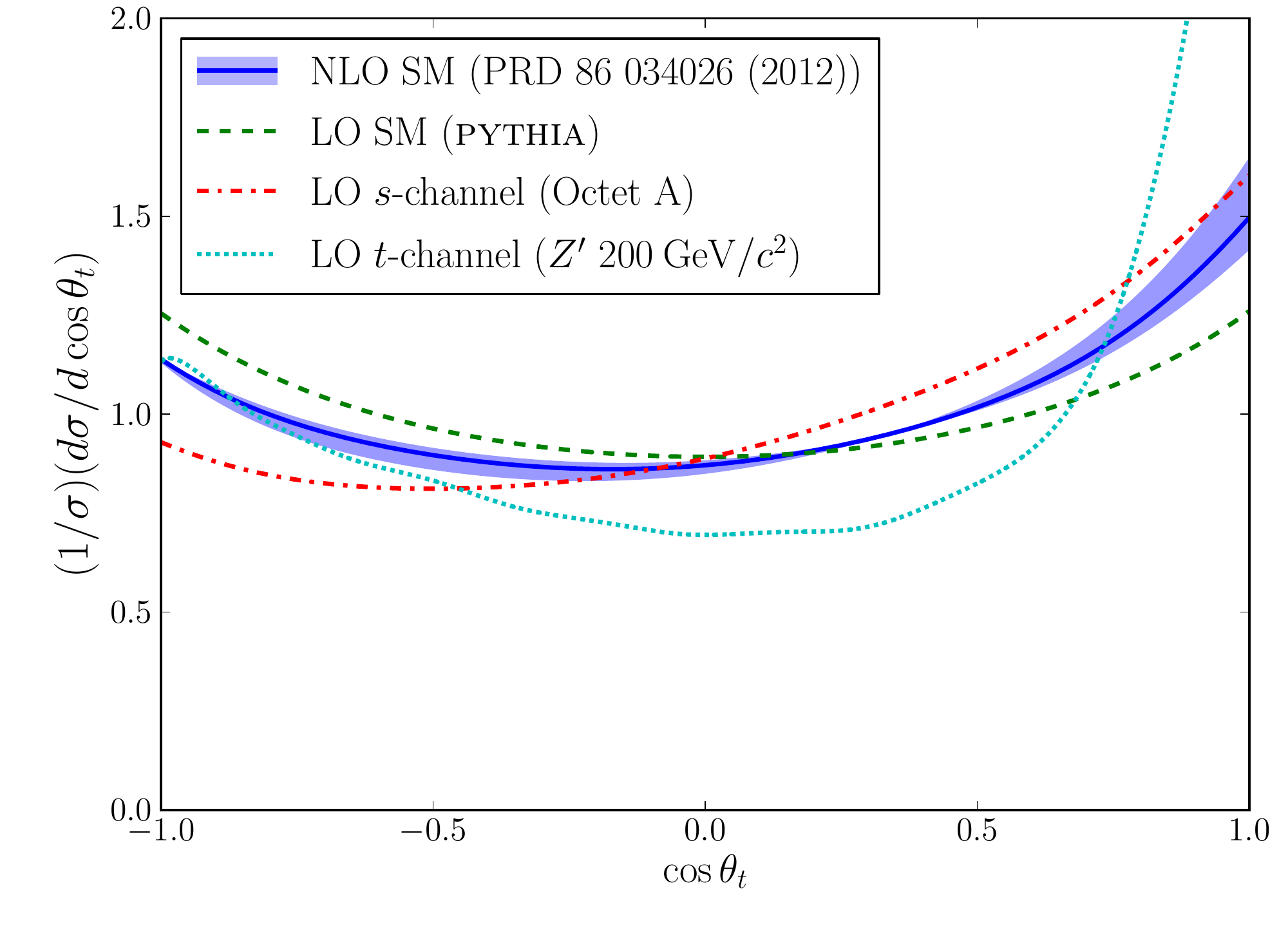}
	\caption{\label{fig:All-True}The predicted differential cross sections of
		the LO SM~\cite{pythia}, NLO SM~\cite{BernSi}, and benchmark models for
		\(s\)- and \(t\)-channel new physics~\cite{CDF2011,JungPearceWells}.
		The band around the NLO SM prediction represents the uncertainty due to
		renormalization-scale choice.}
\end{figure}

At leading order (LO) in the SM, the differential cross section for \(q\bar{q}
\to \ttbar\) is \begin{equation}\frac{d\sigma^{q\bar{q} \to t\bar{t}}}{d\Omega}
\left(\ctt, \hat{s}\right) = \frac{\beta \alpha_s^4}{144 \pi \hat{s}} \left[2 -
\beta^2 (1+\cos^2 \theta_t)\right],\end{equation}where \(\beta\) is the
velocity of the top quark in units of \si{\clight}~\cite{greshamkimzurek} and
\(\hat{s}\) is the Mandelstam variable~\cite{mandelstam}.  After integrating
over \(\hat{s}\) to obtain \dsigctt and comparing to the Legendre polynomials,
we expect non-zero values only for \(a_0\) and \(a_2\).  The addition of the
\(gg \to \ttbar\) process is expected to add small contributions to all the
even-degree Legendre moments.  We study the LO SM via a sample of simulated
events generated by \pythia~\cite{pythia}.  At next-to-leading order in the SM,
additional contributions to all the Legendre moments appear, including the odd
moments.  These non-zero odd moments introduce the lowest-order contributions
to \afb.  The NLO SM theoretical calculation adopted in this Letter includes
the full effects of both quantum chromodynamics and the electroweak
theory~\cite{BernSi}.

\begin{figure}[htbp]
	\includegraphics{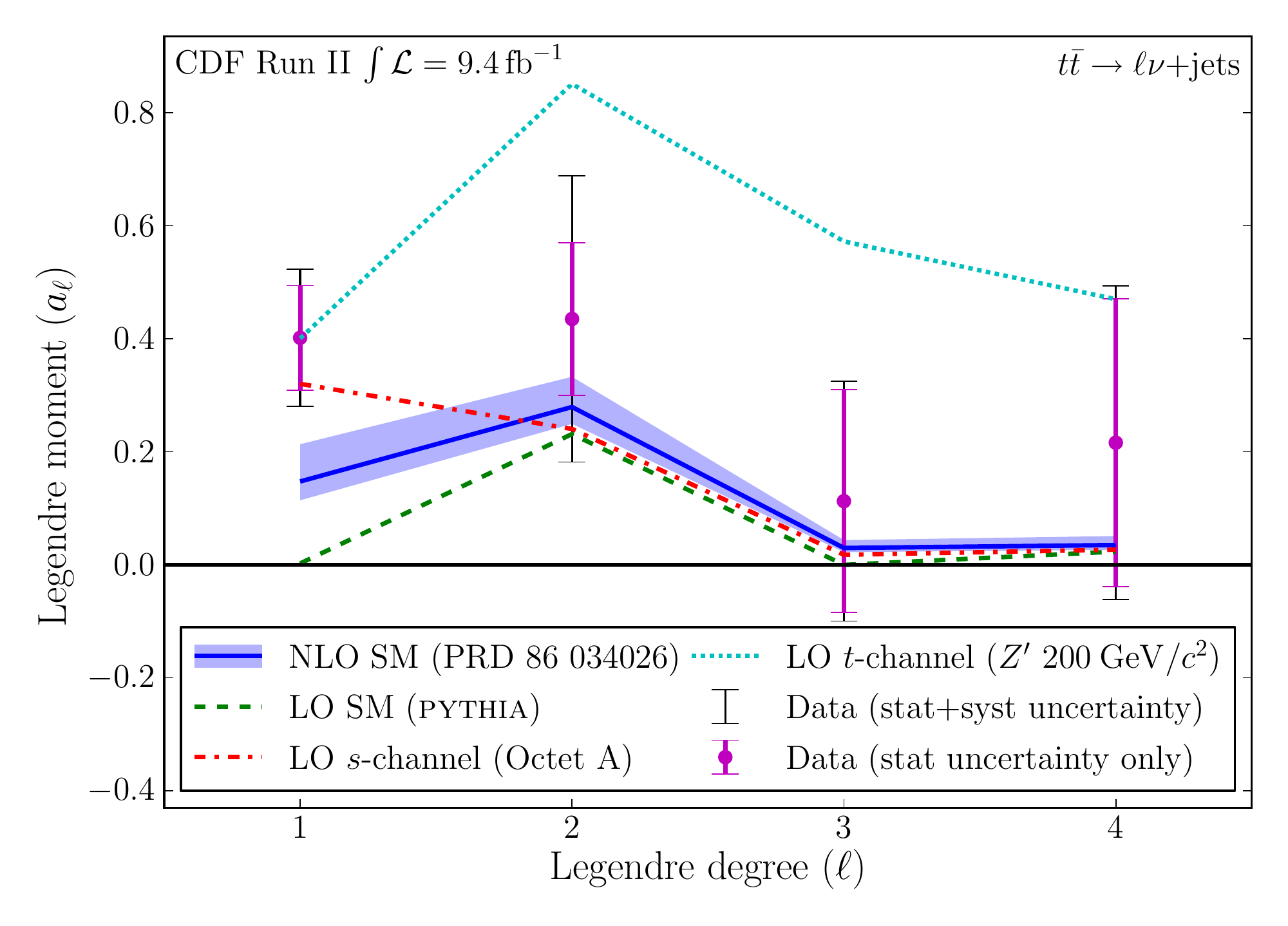}
	\caption{\label{fig:partonmoments}Measured Legendre moments
	\(a_1\)--\(a_4\), with theory predictions overlaid.}
\end{figure}

A wide variety of non-SM proposals has been put forward to explain the large
value of \afb observed at the Tevatron.  These form two broad classes,
depending on whether the new physics is dominated by \(s\)- or \(t\)-channel
exchange.  In order to characterize the effect of these models on the
differential cross section, we study two representative models.  An
\(s\)-channel model, ``Octet~A'', hypothesizes the existence of a heavy
(\(m_{G'} = \SI[per-mode=symbol]{2}{\tera\eV\per\clight\squared}\)) partner of
the gluon with axial-vector couplings to quarks~\cite{CDF2011}.  This produces
an enhanced linear-term coefficient, \(a_1\), in
\dsigctt~\cite{greshamkimzurek}.  A \(t\)-channel model, ``\(Z'\) 200'',
contains a new, heavy (\(m_{Z'} =
\SI[per-mode=symbol]{200}{\giga\eV\per\clight\squared}\)) vector boson with a
flavor changing \(u\)\nobreakdash--\(Z'\)\nobreakdash--\(t\)
coupling~\cite{JungPearceWells}.  The resulting additional term in the cross
section has a leading dependence \(\hat{s} / \hat{t} = 1/(1 - \ctt)\), where
\(\hat{t}\) is the Mandelstam variable~\cite{mandelstam}.  This behavior
produces large Legendre moments at all degrees.  These leading behaviors are
generic predictions of \(s\)- and \(t\)-channel models~\cite{greshamkimzurek}.
Both models are studied via samples of simulated events generated at LO by
\textsc{madgraph}~\cite{Alwall:2007st}.  The LO and NLO SM calculations, as
well as these two benchmark non-SM models, are shown in
Figs.~\ref{fig:All-True} and \ref{fig:partonmoments}.

We study the full sample of top-quark-pair candidate events in the decay
channel with a single lepton in the final state collected by the CDF experiment
during Run II of the Fermilab Tevatron.  The CDF~II detector is a general
purpose particle detector employing a large charged-particle tracking volume
inside a solenoidal magnetic field coaxial with the beam direction, surrounded
by calorimeters and muon detectors~\cite{Acosta:2004yw,geometry}.  The
collected data correspond to an integrated luminosity of
\SI{9.4}{\per\femto\barn} of \ppbar collisions.  The general features of the
event selection requirements are as follows.  We require exactly one
well-reconstructed charged-lepton candidate (electron or muon) with \(p_T >
\SI[per-mode=symbol]{20}{\giga\eV\per\clight}\), an imbalance in the total
event transverse momentum (missing transverse energy~\cite{MET}) \(\MET >
\SI{20}{\giga\eV}\), and four or more calorimeter-energy clusters
(jets~\cite{Bhatti:2005ai}), three with \(E_T > \SI{20}{\giga\eV}\) and the
fourth with \(E_T > \SI{12}{\giga\eV}\), in the central part of the detector
(\(\left\lvert\eta\right\rvert < 2.0\)).  We further require that at least one
of the jets be identified (tagged) as having a displaced vertex resulting from
the decay of a bottom-quark meson, which is produced from the dominant
top-quark decay \(t \to Wb\).  Further details on the online and offline event
selection requirements are in \cite{CDFLJAfb}.  The resulting data set is
enriched in \ttbar events, but it contains non-\ttbar background events as
well, dominated by events in which a \(W\) boson is produced in association
with hadron jets.  The rates and differential distributions of all the sources
of non-\ttbar backgrounds are well understood~\cite{CDFLJAfb}.  We expect to
observe \num{2750(427)} \ttbar events and \num{1026(210)} non-\ttbar background
events, and we observe 3864 \ttbar candidate events.

We reconstruct the top quark and the top anti-quark from their decay products,
using the measured momentum of the lepton and the four jets, as well as the
missing transverse energy.  We fit each possible jet-to-parton assignment to
the \ttbar hypothesis.  We require that two of the jets be consistent with the
decay of a \(W\) boson of mass
\SI[per-mode=symbol]{80.4}{\giga\eV\per\clight\squared} and that the lepton and
missing transverse energy also be consistent with the decay of a \(W\) boson.
We further require that each reconstructed \(W\) boson, when paired with one of
the remaining jets, be consistent with the decay of a top quark of mass
\SI[per-mode=symbol]{172.5}{\giga\eV\per\clight\squared}~\cite{kinfitter}.  The
jet-to-parton assignment which is most consistent with this \ttbar hypothesis
is used to calculate the top-quark production angle as measured in the
detector, \(\ctt^\text{det}\), for each event.

We exploit the orthonormality of the Legendre polynomials to estimate the
Legendre moments without performing a fit.  Given a distribution \(f(\ctt)\),
the Legendre moments of \(f\) are \begin{equation}a_\ell = \frac{2\ell+1}{2}
\int_{-1}^1 d (\ctt) f(\ctt) P_\ell(\ctt).\label{legproj}\end{equation}The data
are described by an empirical distribution~\cite{Kolmogorov,*Smirnov},
\(f(\ctt^\text{det}) = \sum_{i} \delta(\ctt^\text{det} -
\cos\theta^\text{det}_{t,i})\), where \(\delta(x)\) is the Dirac \(\delta\)
function and the index \(i\) runs over the events in the data set.  Using this
distribution in Eq. (\ref{legproj}) greatly simplifies the integration due to
the Dirac delta functions, so the moments of the observed \(\ctt^\text{det}\)
distribution are \begin{equation}a^\text{det}_\ell = \frac{2\ell+1}{2} \sum_i
P_\ell(\cos \theta^\text{det}_{t,i}).\end{equation}Then, the estimate of the
moments is \begin{equation}a_\ell = \sum_m K_{\ell m} \left(\sum_i P_m(\cos
\theta^\text{det}_{t,i}) - a^\text{BG}_m\right),\end{equation}where
\(a^\text{BG}_m\) represents the Legendre moments of the distribution of
\(\ctt^\text{det}\) predicted by the background model, and \(K_{\ell m}\) is a
correction matrix that accounts for the finite resolution of the detector and
for the non-uniform detector acceptance and selection efficiency.  The matrix
\(K\) is developed from a sample of fully-simulated \ttbar Monte Carlo events
generated by the \textsc{powheg} NLO SM generator~\cite{powhegttbar}.  It
describes the response of the detector and the effects of the event selection
requirements.  No smoothing or regularization is applied in this correction
procedure, in contrast to the correction procedure of~\cite{CDFLJAfb}.

The statistical uncertainties on the moments are given by a root-mean-square
covariance matrix including correlations.  In order to estimate the effect from
each of several sources of systematic uncertainty in the model assumptions, we
vary the corresponding nuisance parameter that alters either the background
prediction or correction matrix, and then perform the full correction procedure
again.  The resulting parton-level moments estimate is compared to the unvaried
moments, and then the covariance matrix describing the uncertainty on the
measurement is \(\sigma_{m\ell} = \delta_m \delta_\ell\), where \(\delta_\ell
\equiv a_\ell^\text{varied} - a_\ell^\text{nominal}\).  We study systematic
shifts due to the uncertainty in the jet-energy scale, the rate of the
backgrounds, the shape of the backgrounds, the modeling of parton showering,
the modeling of color reconnection, the modeling of initial- and final-state
radiation, and the parton distribution functions of the proton and antiproton.
We sum the resulting covariance matrices and add them to the statistical
covariance matrix to obtain a covariance matrix that fully describes the
uncertainty of the measurement of the parton-level Legendre moments.  The
eigenvalues and eigenvectors of the covariance matrix~\cite{covarsupp} can be
used to calculate a \(\chi^2\) goodness-of-fit statistic with eight degrees of
freedom in order to perform fits to the data.

\begin{table}[htpb]
	\caption{\label{tab:partonmoments}Measured Legendre moments
	\(a_1\)--\(a_8\), with NLO SM prediction.  The uncertainty on the
	measured moments is the total uncertainty from statistical and
	systematic sources.  The uncertainty on the prediction reflects
	reasonable variations in the renormalization scale~\cite{BernSi}.}
	\sisetup{
		table-figures-decimal=2,
		table-figures-integer=2,
		table-sign-mantissa=true,
		table-figures-uncertainty=2,
		table-figures-exponent=0,
		table-sign-exponent=false,
		tight-spacing=true,
		table-number-alignment=center,
	}
	\begin{tabular}{cSc}
		\hline \hline
		{\(\ell\)} & {\(a_\ell\) (obs)} & {\(a_\ell\) (pred)} \\
		\hline
		1 &  0.40(12) & \(0.15^{+0.07}_{-0.03}\) \\
		2 &  0.44(25) & \(0.28^{+0.05}_{-0.03}\) \\
		3 &  0.11(21) & \(0.030^{+0.014}_{-0.007}\) \\
		4 &  0.22(28) & \(0.035^{+0.016}_{-0.008}\) \\
		5 &  0.11(33) & \(0.005^{+0.002}_{-0.001}\) \\
		6 &  0.24(40) & \(0.006^{+0.002}_{-0.003}\) \\
		7 & -0.15(48) & \(-0.003^{+0.001}_{-0.001}\) \\
		8 &  0.16(65) & \(-0.0019^{+0.0003}_{-0.0003}\) \\
		\hline \hline
	\end{tabular}
\end{table}

The parton-level Legendre moments are shown in \figref{fig:partonmoments} and
in \tabref{tab:partonmoments}.  We observe good agreement within the
uncertainties with the NLO SM prediction for moments
\(a_2\)\nobreakdash--\(a_8\), but \(a_1\) is in excess of the prediction.  That
is, a mild excess is observed in the differential cross section in the term
linear in \ctt, while all other terms are as predicted by the SM.  The LO SM
prediction is strongly disfavored by the linear term, with a significance of
more than three standard deviations.  The benchmark \(t\)-channel model,
``\(Z'\) 200'', is disfavored by \(a_2\) and \(a_3\).  The benchmark
\(s\)-channel model, ``Octet~A'', is in good agreement with the data.

\begin{figure}[htpb]
	\includegraphics{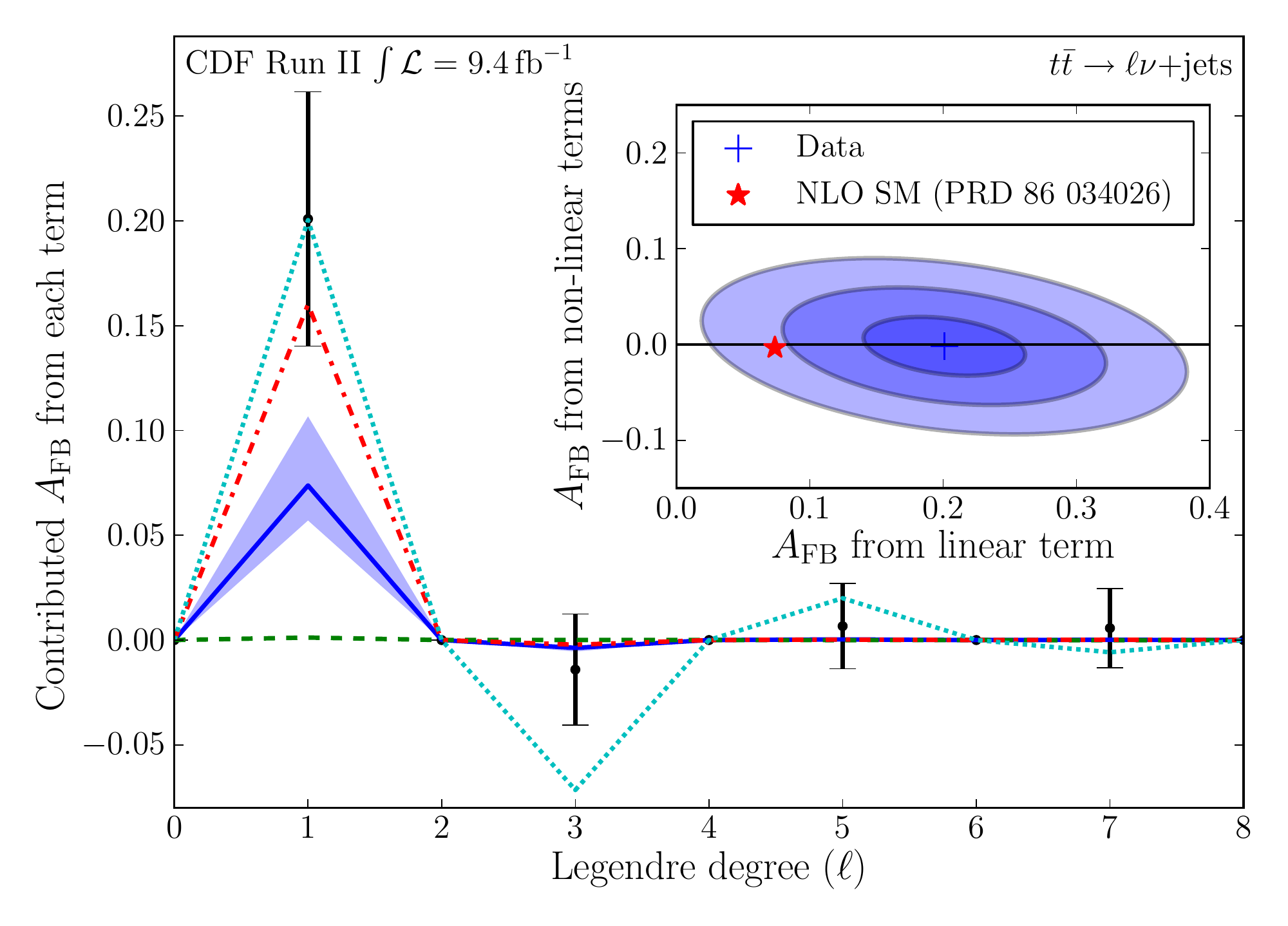}
	\caption{\label{fig:AFB}Absolute contributions of the Legendre moments to
	the \afb, with theory predictions overlaid.  The lines and symbols are the
	same as in \figref{fig:partonmoments}.  The inset shows the 1-, 2-, and
	3-standard-deviation uncertainty ellipses.}
\end{figure}

We determine the contribution of each Legendre moment to the \afb from the
inherent asymmetry of each polynomial (\figref{fig:AFB}).  The observed \afb
\SI{19.9(57)}{\percent} is completely dominated by the excess linear term,
\(a_1 \ctt\), which contributes \SI{20.1(61)}{\percent}.  The \afb contributed
by the non-linear asymmetric terms \(a_3\), \(a_5\), and \(a_7\) is negligible
\SI{-0.2(31)}{\percent}, and is consistent with the SM prediction
(\SI{7.3}{\percent} from the linear term, \SI{-0.3}{\percent} from the
non-linear terms).  The correlation between the measurements of \afb from the
linear and non-linear terms is \SI{-29}{\percent}.

\begin{figure}[htpb]
	\includegraphics{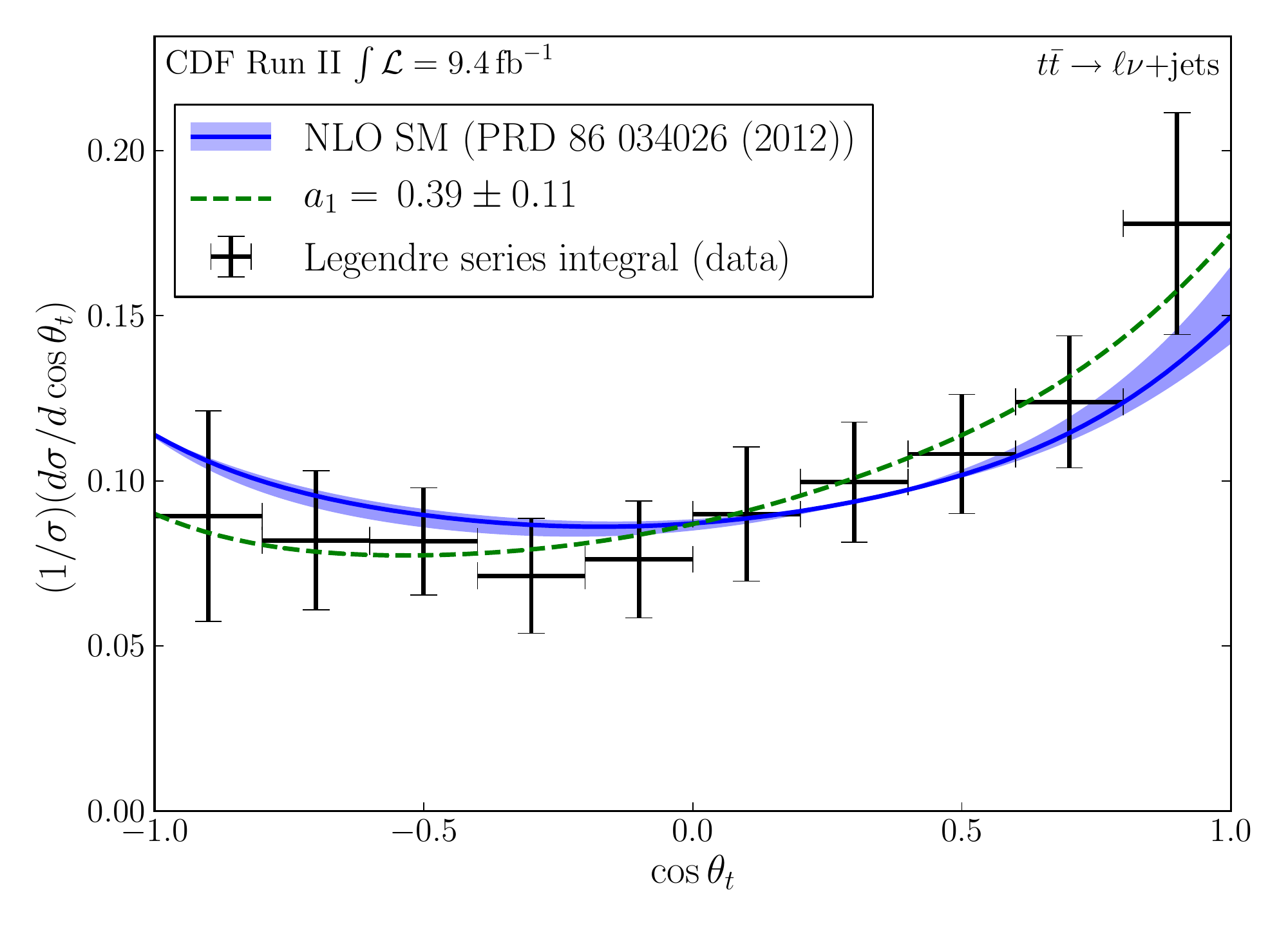}
	\caption{\label{fig:Binned}Fraction of cross section accruing in 10 bins of
	\ctt, obtained by integrating the series of Legendre polynomials over the
	width of each bin.}
\end{figure}

A more traditional picture of the differential cross section
(\figref{fig:Binned}) is obtained by integrating the Legendre series over
 intervals (bins) in \ctt.  This shows the fraction of the total cross
section that accrues in each bin.  The uncertainties are strongly correlated,
and they are dominated by the large uncertainties on the high-degree Legendre
moments.

Because the non-linear moments, \(a_2\)\nobreakdash--\(a_8\), are in good
agreement within the uncertainties with the NLO SM prediction, we may obtain a
more precise, but model-dependent, estimate of the linear term by explicitly
assuming that the non-linear moments are as predicted by the NLO SM
calculation.  Using the covariance matrix and the fitting procedure described
in~\cite{covarsupp}, we fit to the measured moments, taking the NLO SM
prediction for the non-linear moments with their scale uncertainties as a prior
assumption, obtaining \(a_1 = \num{0.39(11)}\) (including statistical and
systematic uncertainty).  Through the correlations among the measured moments,
this reduces the uncertainty on \(a_1\) by about \SI{10}{\percent} while
shifting the central value less than \SI{3}{\percent}.  The resulting curve is
also shown in \figref{fig:Binned}.

In conclusion, we have presented the first measurement of the top-quark-pair
production differential cross section, \dsigctt, in \ppbar collisions at
\(\sqrt{s} = \SI{1.96}{\tera\eV}\) as a function of the production angle of the
top quark.  In order to probe the origin of the top-quark-production asymmetry,
we decompose the angular form into Legendre polynomials.  We observe that the
coefficient of the \ctt term in the differential cross section, \(a_1 =
\num{0.40(12)}\), is in excess of the NLO SM prediction,
\(0.15^{+0.07}_{-0.03}\), while the remainder of the differential cross section
is in good agreement within the uncertainties with the NLO SM prediction.  The
top-quark forward-backward asymmetry is thus completely dominated by the linear
term.  The result constrains \(t\)-channel explanations of the asymmetry and
favors asymmetry models with strong \(s\)-channel components.

\begin{acknowledgments}
We thank T.\ Tait, S.\ Jung, W.\ Bernreuther, and Z.-G.\ Si for their
assistance in preparing the theoretical models and calculations used in this
Letter, and T.\ Rizzo for helpful conversations.  We also thank the development
teams of \textsc{scipy}, \textsc{pytables}, \textsc{matplotlib}, and
\textsc{ipython} for their useful
tools~\cite{scipy,*pytables,*matplotlib,*ipython}.

We thank the Fermilab staff and the technical staffs of the participating
institutions for their vital contributions. This work was supported by the U.S.
Department of Energy and National Science Foundation; the Italian Istituto
Nazionale di Fisica Nucleare; the Ministry of Education, Culture, Sports,
Science and Technology of Japan; the Natural Sciences and Engineering Research
Council of Canada; the National Science Council of the Republic of China; the
Swiss National Science Foundation; the A.P. Sloan Foundation; the
Bundesministerium f\"ur Bildung und Forschung, Germany; the Korean World Class
University Program, the National Research Foundation of Korea; the Science and
Technology Facilities Council and the Royal Society, UK;  the Russian
Foundation for Basic Research; the Ministerio de Ciencia e Innovaci\'{o}n, and
Programa Consolider-Ingenio 2010, Spain; the Slovak R\&D Agency; the Academy of
Finland; the Australian Research Council (ARC); and the EU community Marie
Curie Fellowship contract 302103. 

\end{acknowledgments}

\bibliographystyle{apsrev4-1-JHEPfix}
\bibliography{cosPRL}

\begin{thebibliography}{38}%
\makeatletter
\providecommand \@ifxundefined [1]{%
 \@ifx{#1\undefined}
}%
\providecommand \@ifnum [1]{%
 \ifnum #1\expandafter \@firstoftwo
 \else \expandafter \@secondoftwo
 \fi
}%
\providecommand \@ifx [1]{%
 \ifx #1\expandafter \@firstoftwo
 \else \expandafter \@secondoftwo
 \fi
}%
\providecommand \natexlab [1]{#1}%
\providecommand \enquote  [1]{``#1''}%
\providecommand \bibnamefont  [1]{#1}%
\providecommand \bibfnamefont [1]{#1}%
\providecommand \citenamefont [1]{#1}%
\providecommand \href@noop [0]{\@secondoftwo}%
\providecommand \href [0]{\begingroup \@sanitize@url \@href}%
\providecommand \@href[1]{\@@startlink{#1}\@@href}%
\providecommand \@@href[1]{\endgroup#1\@@endlink}%
\providecommand \@sanitize@url [0]{\catcode `\\12\catcode `\$12\catcode
  `\&12\catcode `\#12\catcode `\^12\catcode `\_12\catcode `\%12\relax}%
\providecommand \@@startlink[1]{}%
\providecommand \@@endlink[0]{}%
\providecommand \url  [0]{\begingroup\@sanitize@url \@url }%
\providecommand \@url [1]{\endgroup\@href {#1}{\urlprefix }}%
\providecommand \urlprefix  [0]{URL }%
\providecommand \Eprint [0]{\href }%
\providecommand \doibase [0]{http://dx.doi.org/}%
\providecommand \selectlanguage [0]{\@gobble}%
\providecommand \bibinfo  [0]{\@secondoftwo}%
\providecommand \bibfield  [0]{\@secondoftwo}%
\providecommand \translation [1]{[#1]}%
\providecommand \BibitemOpen [0]{}%
\providecommand \bibitemStop [0]{}%
\providecommand \bibitemNoStop [0]{.\EOS\space}%
\providecommand \EOS [0]{\spacefactor3000\relax}%
\providecommand \BibitemShut  [1]{\csname bibitem#1\endcsname}%
\let\auto@bib@innerbib\@empty
\bibitem [{\citenamefont {Aaltonen}\ \emph {et~al.}(2013)\citenamefont
  {Aaltonen} \emph {et~al.}}]{CDFLJAfb}%
  \BibitemOpen
  \bibfield  {author} {\bibinfo {author} {\bibfnamefont {T.}~\bibnamefont
  {Aaltonen}}\  \emph {et~al.} (\bibinfo {collaboration} {CDF Collaboration}),\
  }\href {\doibase 10.1103/PhysRevD.87.092002} {\bibfield  {journal} {\bibinfo
  {journal} {Phys. Rev. D}\ }\textbf {\bibinfo {volume} {87}},\ \bibinfo
  {pages} {092002} (\bibinfo {year} {2013})}\BibitemShut {NoStop}%
\bibitem [{\citenamefont {Abazov}\ \emph {et~al.}(2011)\citenamefont {Abazov}
  \emph {et~al.}}]{D0LJAfb}%
  \BibitemOpen
  \bibfield  {author} {\bibinfo {author} {\bibfnamefont {V.~M.}\ \bibnamefont
  {Abazov}}\  \emph {et~al.} (\bibinfo {collaboration} {D0 Collaboration}),\
  }\href {\doibase 10.1103/PhysRevD.84.112005} {\bibfield  {journal} {\bibinfo
  {journal} {Phys. Rev. D}\ }\textbf {\bibinfo {volume} {84}},\ \bibinfo
  {pages} {112005} (\bibinfo {year} {2011})}\BibitemShut {NoStop}%
\bibitem [{Note1()}]{Note1}%
  \BibitemOpen
  \bibinfo {note} {\protect \textit {CP} is the simultaneous transformation of
  charge conjugation and parity.}\BibitemShut {Stop}%
\bibitem [{\citenamefont {Glashow}(1961)}]{Glashow:1961tr}%
  \BibitemOpen
  \bibfield  {author} {\bibinfo {author} {\bibfnamefont {S.~L.}\ \bibnamefont
  {Glashow}},\ }\href {\doibase 10.1016/0029-5582(61)90469-2} {\bibfield
  {journal} {\bibinfo  {journal} {Nucl. Phys.}\ }\textbf {\bibinfo {volume}
  {22}},\ \bibinfo {pages} {579} (\bibinfo {year} {1961})}\BibitemShut
  {NoStop}%
\bibitem [{\citenamefont {Weinberg}(1967)}]{Weinberg:1967tq}%
  \BibitemOpen
  \bibfield  {author} {\bibinfo {author} {\bibfnamefont {S.}~\bibnamefont
  {Weinberg}},\ }\href {\doibase 10.1103/PhysRevLett.19.1264} {\bibfield
  {journal} {\bibinfo  {journal} {Phys. Rev. Lett.}\ }\textbf {\bibinfo
  {volume} {19}},\ \bibinfo {pages} {1264} (\bibinfo {year}
  {1967})}\BibitemShut {NoStop}%
\bibitem [{\citenamefont {Salam}(1968)}]{Salam:1968}%
  \BibitemOpen
  \bibfield  {author} {\bibinfo {author} {\bibfnamefont {A.}~\bibnamefont
  {Salam}},\ }in\ \href@noop {} {\emph {\bibinfo {booktitle} {Elementary
  Particle Theory}}},\ \bibinfo {editor} {edited by\ \bibinfo {editor}
  {\bibfnamefont {N.}~\bibnamefont {Svartholm}}}\ (\bibinfo  {publisher}
  {Almquist and Wiksell, Stockholm},\ \bibinfo {year} {1968})\ p.\ \bibinfo
  {pages} {367}\BibitemShut {NoStop}%
\bibitem [{\citenamefont {Bernreuther}\ and\ \citenamefont
  {Si}(2012)}]{BernSi}%
  \BibitemOpen
  \bibfield  {author} {\bibinfo {author} {\bibfnamefont {W.}~\bibnamefont
  {Bernreuther}}\ and\ \bibinfo {author} {\bibfnamefont {Z.-G.}\ \bibnamefont
  {Si}},\ }\href {\doibase 10.1103/PhysRevD.86.034026} {\bibfield  {journal}
  {\bibinfo  {journal} {Phys. Rev. D}\ }\textbf {\bibinfo {volume} {86}},\
  \bibinfo {pages} {034026} (\bibinfo {year} {2012})}\BibitemShut {NoStop}%
\bibitem [{Ber({\natexlab{a}})}]{BernPriv}%
  \BibitemOpen
  \bibinfo {note} {W.~Bernreuther (private communication)}\BibitemShut
  {NoStop}%
\bibitem [{\citenamefont {Kuhn}\ and\ \citenamefont
  {Rodrigo}(1998)}]{KuhnRodrigo98}%
  \BibitemOpen
  \bibfield  {author} {\bibinfo {author} {\bibfnamefont {J.~H.}\ \bibnamefont
  {Kuhn}}\ and\ \bibinfo {author} {\bibfnamefont {G.}~\bibnamefont {Rodrigo}},\
  }\href {\doibase 10.1103/PhysRevLett.81.49} {\bibfield  {journal} {\bibinfo
  {journal} {Phys. Rev. Lett.}\ }\textbf {\bibinfo {volume} {81}},\ \bibinfo
  {pages} {49} (\bibinfo {year} {1998})}\BibitemShut {NoStop}%
\bibitem [{\citenamefont {Brodsky}\ and\ \citenamefont {Wu}(2012)}]{Brodsky}%
  \BibitemOpen
  \bibfield  {author} {\bibinfo {author} {\bibfnamefont {S.~J.}\ \bibnamefont
  {Brodsky}}\ and\ \bibinfo {author} {\bibfnamefont {X.-G.}\ \bibnamefont
  {Wu}},\ }\href {\doibase 10.1103/PhysRevD.85.114040} {\bibfield  {journal}
  {\bibinfo  {journal} {Phys. Rev. D}\ }\textbf {\bibinfo {volume} {85}},\
  \bibinfo {pages} {114040} (\bibinfo {year} {2012})}\BibitemShut {NoStop}%
\bibitem [{\citenamefont {Hollik}\ and\ \citenamefont
  {Pagani}(2011)}]{HollikEWK}%
  \BibitemOpen
  \bibfield  {author} {\bibinfo {author} {\bibfnamefont {W.}~\bibnamefont
  {Hollik}}\ and\ \bibinfo {author} {\bibfnamefont {D.}~\bibnamefont
  {Pagani}},\ }\href {\doibase 10.1007/JHEP01(2012)063} {\bibfield  {journal}
  {\bibinfo  {journal} {Phys. Rev. D}\ }\textbf {\bibinfo {volume} {84}},\
  \bibinfo {pages} {093003} (\bibinfo {year} {2011})}\BibitemShut {NoStop}%
\bibitem [{\citenamefont {Kuhn}\ and\ \citenamefont {Rodrigo}(2012)}]{KuhnEWK}%
  \BibitemOpen
  \bibfield  {author} {\bibinfo {author} {\bibfnamefont {J.~H.}\ \bibnamefont
  {Kuhn}}\ and\ \bibinfo {author} {\bibfnamefont {G.}~\bibnamefont {Rodrigo}},\
  }\href {\doibase 10.1007/JHEP01(2012)063} {\bibfield  {journal} {\bibinfo
  {journal} {J. High Energy Phys.}\ }\bibinfo {volume} {01} (\bibinfo {year}
  {2012})\ \bibinfo {pages} {063}}\BibitemShut {NoStop}%
\bibitem [{\citenamefont {Manohar}\ and\ \citenamefont
  {Trott}(2012)}]{ManoharEWK}%
  \BibitemOpen
  \bibfield  {author} {\bibinfo {author} {\bibfnamefont {A.~V.}\ \bibnamefont
  {Manohar}}\ and\ \bibinfo {author} {\bibfnamefont {M.}~\bibnamefont
  {Trott}},\ }\href {\doibase 10.1016/j.physletb.2012.04.013} {\bibfield
  {journal} {\bibinfo  {journal} {Phys. Lett. B}\ }\textbf {\bibinfo {volume}
  {711}},\ \bibinfo {pages} {313} (\bibinfo {year} {2012})}\BibitemShut
  {NoStop}%
\bibitem [{\citenamefont {Kamenik}\ \emph {et~al.}(2012)\citenamefont
  {Kamenik}, \citenamefont {Shu},\ and\ \citenamefont {Zupan}}]{BSMReviews}%
  \BibitemOpen
  \bibfield  {author} {\bibinfo {author} {\bibfnamefont {J.~F.}\ \bibnamefont
  {Kamenik}}, \bibinfo {author} {\bibfnamefont {J.}~\bibnamefont {Shu}}, and\
  \bibinfo {author} {\bibfnamefont {J.}~\bibnamefont {Zupan}},\ }\href
  {\doibase 10.1140/epjc/s10052-012-2102-2} {\bibfield  {journal} {\bibinfo
  {journal} {Eur. Phys. J. C}\ }\textbf {\bibinfo {volume} {72}},\ \bibinfo
  {pages} {2102} (\bibinfo {year} {2012})}\BibitemShut {NoStop}%
\bibitem [{Ber({\natexlab{b}})}]{Berger}%
  \BibitemOpen
  \bibinfo {note} {E.~L.~Berger, Argonne National Laboratory Report
  ANL-HEP-CP-12-87, 2013 (unpublished).}\BibitemShut {Stop}%
\bibitem [{\citenamefont {Chatrchyan}\ \emph {et~al.}(2012)\citenamefont
  {Chatrchyan} \emph {et~al.}}]{CMSAfb}%
  \BibitemOpen
  \bibfield  {author} {\bibinfo {author} {\bibfnamefont {S.}~\bibnamefont
  {Chatrchyan}}\  \emph {et~al.} (\bibinfo {collaboration} {CMS
  Collaboration}),\ }\href {\doibase 10.1016/j.physletb.2012.09.028} {\bibfield
   {journal} {\bibinfo  {journal} {Phys. Lett. B}\ }\textbf {\bibinfo {volume}
  {717}},\ \bibinfo {pages} {129} (\bibinfo {year} {2012})}\BibitemShut
  {NoStop}%
\bibitem [{\citenamefont {Aad}\ \emph {et~al.}(2012)\citenamefont {Aad} \emph
  {et~al.}}]{ATLASAfb}%
  \BibitemOpen
  \bibfield  {author} {\bibinfo {author} {\bibfnamefont {G.}~\bibnamefont
  {Aad}}\  \emph {et~al.} (\bibinfo {collaboration} {{ATLAS Collaboration}}),\
  }\href {\doibase 10.1140/epjc/s10052-012-2039-5} {\bibfield  {journal}
  {\bibinfo  {journal} {Euro. Phys. J. C}\ }\textbf {\bibinfo {volume} {72}},\
  \bibinfo {pages} {2039} (\bibinfo {year} {2012})}\BibitemShut {NoStop}%
\bibitem [{Note2()}]{Note2}%
  \BibitemOpen
  \bibinfo {note} {We normalize the Legendre polynomials so that \(\protect
  \frac {2\ell +1}{2} \DOTSI \intop \ilimits@ _{-1}^{1} d x P_{\ell }(x)
  P_{\ell '}(x) = \delta _{\ell \ell '}\), where \(\delta _{\ell \ell '}\) is
  the Kronecker delta.}\BibitemShut {Stop}%
\bibitem [{jac()}]{jacobwick}%
  \BibitemOpen
  \bibinfo {note} {M.~Jacob and G.C.~Wick, Ann. Phys. (N.Y.) {\bf 7}, 404
  (1959); {\bf 281}, 774 (2000).}\BibitemShut {Stop}%
\bibitem [{\citenamefont {Sj{\"o}strand}\ \emph {et~al.}(2006)\citenamefont
  {Sj{\"o}strand}, \citenamefont {Mrenna},\ and\ \citenamefont
  {Skands}}]{pythia}%
  \BibitemOpen
  \bibfield  {author} {\bibinfo {author} {\bibfnamefont {T.}~\bibnamefont
  {Sj{\"o}strand}}, \bibinfo {author} {\bibfnamefont {S.}~\bibnamefont
  {Mrenna}}, and\ \bibinfo {author} {\bibfnamefont {P.}~\bibnamefont
  {Skands}},\ }\href {\doibase 10.1088/1126-6708/2006/05/026} {\bibfield
  {journal} {\bibinfo  {journal} {J. High Energy Phys.}\ }\bibinfo {volume}
  {05} (\bibinfo {year} {2006})\ \bibinfo {pages} {026}},\ \bibinfo {note} {we
  use \pythia\ version 6.216.}\BibitemShut {Stop}%
\bibitem [{\citenamefont {Aaltonen}\ \emph {et~al.}(2011)\citenamefont
  {Aaltonen} \emph {et~al.}}]{CDF2011}%
  \BibitemOpen
  \bibfield  {author} {\bibinfo {author} {\bibfnamefont {T.}~\bibnamefont
  {Aaltonen}}\  \emph {et~al.} (\bibinfo {collaboration} {CDF Collaboration}),\
  }\href {\doibase 10.1103/PhysRevD.85.114040} {\bibfield  {journal} {\bibinfo
  {journal} {Phys. Rev. D}\ }\textbf {\bibinfo {volume} {83}},\ \bibinfo
  {pages} {112003} (\bibinfo {year} {2011})}\BibitemShut {NoStop}%
\bibitem [{\citenamefont {Jung}\ \emph {et~al.}(2011)\citenamefont {Jung},
  \citenamefont {Pierce},\ and\ \citenamefont {Wells}}]{JungPearceWells}%
  \BibitemOpen
  \bibfield  {author} {\bibinfo {author} {\bibfnamefont {S.}~\bibnamefont
  {Jung}}, \bibinfo {author} {\bibfnamefont {A.}~\bibnamefont {Pierce}}, and\
  \bibinfo {author} {\bibfnamefont {J.~D.}\ \bibnamefont {Wells}},\ }\href
  {\doibase 10.1103/PhysRevD.83.114039} {\bibfield  {journal} {\bibinfo
  {journal} {Phys. Rev. D}\ }\textbf {\bibinfo {volume} {83}},\ \bibinfo
  {pages} {114039} (\bibinfo {year} {2011})}\BibitemShut {NoStop}%
\bibitem [{\citenamefont {Gresham}\ \emph {et~al.}(2011)\citenamefont
  {Gresham}, \citenamefont {Kim},\ and\ \citenamefont
  {Zurek}}]{greshamkimzurek}%
  \BibitemOpen
  \bibfield  {author} {\bibinfo {author} {\bibfnamefont {M.~I.}\ \bibnamefont
  {Gresham}}, \bibinfo {author} {\bibfnamefont {I.-W.}\ \bibnamefont {Kim}},
  and\ \bibinfo {author} {\bibfnamefont {K.~M.}\ \bibnamefont {Zurek}},\ }\href
  {\doibase 10.1103/PhysRevD.83.114027} {\bibfield  {journal} {\bibinfo
  {journal} {Phys. Rev. D}\ }\textbf {\bibinfo {volume} {83}},\ \bibinfo
  {pages} {114027} (\bibinfo {year} {2011})}\BibitemShut {NoStop}%
\bibitem [{\citenamefont {Mandelstam}(1958)}]{mandelstam}%
  \BibitemOpen
  \bibfield  {author} {\bibinfo {author} {\bibfnamefont {S.}~\bibnamefont
  {Mandelstam}},\ }\href {\doibase 10.1103/PhysRev.112.1344} {\bibfield
  {journal} {\bibinfo  {journal} {Phys. Rev.}\ }\textbf {\bibinfo {volume}
  {112}},\ \bibinfo {pages} {1344} (\bibinfo {year} {1958})}\BibitemShut
  {NoStop}%
\bibitem [{\citenamefont {Alwall}\ \emph {et~al.}(2007)\citenamefont {Alwall},
  \citenamefont {Demin}, \citenamefont {de~Visscher}, \citenamefont {Frederix},
  \citenamefont {Herquet}, \citenamefont {Maltoni}, \citenamefont {Plehn},
  \citenamefont {Rainwater},\ and\ \citenamefont {Stelzer}}]{Alwall:2007st}%
  \BibitemOpen
  \bibfield  {author} {\bibinfo {author} {\bibfnamefont {J.}~\bibnamefont
  {Alwall}}, \bibinfo {author} {\bibfnamefont {P.}~\bibnamefont {Demin}},
  \bibinfo {author} {\bibfnamefont {S.}~\bibnamefont {de~Visscher}}, \bibinfo
  {author} {\bibfnamefont {R.}~\bibnamefont {Frederix}}, \bibinfo {author}
  {\bibfnamefont {M.}~\bibnamefont {Herquet}}, \bibinfo {author} {\bibfnamefont
  {F.}~\bibnamefont {Maltoni}}, \bibinfo {author} {\bibfnamefont
  {T.}~\bibnamefont {Plehn}}, \bibinfo {author} {\bibfnamefont {D.~L.}\
  \bibnamefont {Rainwater}}, and\ \bibinfo {author} {\bibfnamefont
  {T.}~\bibnamefont {Stelzer}},\ }\href {\doibase
  10.1088/1126-6708/2007/09/028} {\bibfield  {journal} {\bibinfo  {journal} {J.
  High Energy Phys.}\ }\bibinfo {volume} {09} (\bibinfo {year} {2007})\
  \bibinfo {pages} {028}}\BibitemShut {NoStop}%
\bibitem [{\citenamefont {Acosta}\ \emph {et~al.}(2005)\citenamefont {Acosta}
  \emph {et~al.}}]{Acosta:2004yw}%
  \BibitemOpen
  \bibfield  {author} {\bibinfo {author} {\bibfnamefont {D.~E.}\ \bibnamefont
  {Acosta}}\  \emph {et~al.} (\bibinfo {collaboration} {CDF Collaboration}),\
  }\href {\doibase 10.1103/PhysRevD.71.032001} {\bibfield  {journal} {\bibinfo
  {journal} {Phys. Rev. D}\ }\textbf {\bibinfo {volume} {71}},\ \bibinfo
  {pages} {032001} (\bibinfo {year} {2005})}\BibitemShut {NoStop}%
\bibitem [{geo()}]{geometry}%
  \BibitemOpen
  \bibinfo {note} {We use a cylindrical coordinate system with the origin at
  the center of the CDF detector, $z$ pointing in the direction of the proton
  beam, $\theta$ and $\phi$ representing the polar and azimuthal angles,
  respectively, and pseudorapidity defined by $\eta = - \rm ln$ $\rm
  tan(\theta/2)$. The transverse momentum $p_T$ (transverse energy $E_{T}$) is
  defined to be $p\sin\theta$ ($E\sin\theta$).}\BibitemShut {Stop}%
\bibitem [{MET()}]{MET}%
  \BibitemOpen
  \bibinfo {note} {The calorimeter missing $E_T$ ($\METVEC(\mbox{cal})$) is
  defined by the sum over calorimeter cells, $\METVEC(\mbox{cal}) = - \sum_{i}
  E_T^i \hat{n}_i$, where $i$ is calorimeter cell number with $|\eta| < 3.6$,
  $\hat{n}_i$ is a unit vector perpendicular to the beam axis and pointing at
  the $i$th calorimeter cell. The reconstructed missing energy, $\METVEC$, is
  derived by subtracting from $\METVEC(\mbox{cal})$ the energies associated
  with components of the event not registered by the calorimeter, such as muons
  and jet energy adjustments. $\MET(\mbox{cal})$ and $\MET$ are the scalar
  magnitudes of $\METVEC(\mbox{cal})$ and $\METVEC$, respectively.}\BibitemShut
  {Stop}%
\bibitem [{\citenamefont {Bhatti}\ \emph {et~al.}(2006)\citenamefont {Bhatti}
  \emph {et~al.}}]{Bhatti:2005ai}%
  \BibitemOpen
  \bibfield  {author} {\bibinfo {author} {\bibfnamefont {A.}~\bibnamefont
  {Bhatti}}\  \emph {et~al.},\ }\href {\doibase 10.1016/j.nima.2006.05.269}
  {\bibfield  {journal} {\bibinfo  {journal} {Nucl. Instrum. Methods A}\
  }\textbf {\bibinfo {volume} {566}},\ \bibinfo {pages} {375} (\bibinfo {year}
  {2006})}\BibitemShut {NoStop}%
\bibitem [{\citenamefont {Abulencia}\ \emph {et~al.}(2006)\citenamefont
  {Abulencia} \emph {et~al.}}]{kinfitter}%
  \BibitemOpen
  \bibfield  {author} {\bibinfo {author} {\bibfnamefont {A.}~\bibnamefont
  {Abulencia}}\  \emph {et~al.} (\bibinfo {collaboration} {CDF
  Collaboration}),\ }\href {\doibase 10.1103/PhysRevD.73.032003} {\bibfield
  {journal} {\bibinfo  {journal} {Phys. Rev. D}\ }\textbf {\bibinfo {volume}
  {73}},\ \bibinfo {pages} {032003} (\bibinfo {year} {2006})}\BibitemShut
  {NoStop}%
\bibitem [{\citenamefont {Kolmogorov}(1933)}]{Kolmogorov}%
  \BibitemOpen
  \bibfield  {author} {\bibinfo {author} {\bibfnamefont {A.}~\bibnamefont
  {Kolmogorov}},\ }\href@noop {} {\bibfield  {journal} {\bibinfo  {journal} {G.
  Ist. Ital. Attuari}\ }\textbf {\bibinfo {volume} {4}},\ \bibinfo {pages} {83}
  (\bibinfo {year} {1933})}\BibitemShut {NoStop}%
\bibitem [{\citenamefont {Smirnov}(1938)}]{Smirnov}%
  \BibitemOpen
  \bibfield  {author} {\bibinfo {author} {\bibfnamefont {V.~I.}\ \bibnamefont
  {Smirnov}},\ }\href@noop {} {\bibfield  {journal} {\bibinfo  {journal}
  {Byull. Moskov. Gos. Univ. Ser. A}\ }\textbf {\bibinfo {volume} {2}},\
  \bibinfo {pages} {2} (\bibinfo {year} {1938})}\BibitemShut {NoStop}%
\bibitem [{\citenamefont {Frixione}\ \emph {et~al.}(2007)\citenamefont
  {Frixione}, \citenamefont {Nason},\ and\ \citenamefont
  {Ridolfi}}]{powhegttbar}%
  \BibitemOpen
  \bibfield  {author} {\bibinfo {author} {\bibfnamefont {S.}~\bibnamefont
  {Frixione}}, \bibinfo {author} {\bibfnamefont {P.}~\bibnamefont {Nason}},
  and\ \bibinfo {author} {\bibfnamefont {G.}~\bibnamefont {Ridolfi}},\ }\href
  {\doibase 10.1103/PhysRevD.85.114040} {\bibfield  {journal} {\bibinfo
  {journal} {J. High Energy Phys.}\ }\bibinfo {volume} {09} (\bibinfo {year}
  {2007})\ \bibinfo {pages} {126}}\BibitemShut {NoStop}%
\bibitem [{cov()}]{covarsupp}%
  \BibitemOpen
  \bibinfo {note} {See Supplemental Material at [URL will be inserted by
  publisher] for the eigenvalues and eigenvectors of the covariance matrix
  describing the correlations and uncertainty of this measurement.}\BibitemShut
  {Stop}%
\bibitem [{\citenamefont {Jones}\ \emph {et~al.}(2001)\citenamefont {Jones}
  \emph {et~al.}}]{scipy}%
  \BibitemOpen
  \bibfield  {author} {\bibinfo {author} {\bibfnamefont {E.}~\bibnamefont
  {Jones}}\  \emph {et~al.},\ }\href@noop {} {\enquote {\bibinfo {title}
  {{SciPy}: Open source scientific tools for {Python}},}\ } (\bibinfo {year}
  {2001}),\ \bibinfo {note} {\url{http://www.scipy.org/}}\BibitemShut {NoStop}%
\bibitem [{\citenamefont {Alted}\ \emph {et~al.}(2002)\citenamefont {Alted}
  \emph {et~al.}}]{pytables}%
  \BibitemOpen
  \bibfield  {author} {\bibinfo {author} {\bibfnamefont {F.}~\bibnamefont
  {Alted}}\  \emph {et~al.},\ }\href@noop {} {\enquote {\bibinfo {title}
  {{PyTables}: Hierarchical datasets in {Python}},}\ } (\bibinfo {year}
  {2002}),\ \bibinfo {note} {\url{http://www.pytables.org/}}\BibitemShut
  {NoStop}%
\bibitem [{\citenamefont {Hunter}(2007)}]{matplotlib}%
  \BibitemOpen
  \bibfield  {author} {\bibinfo {author} {\bibfnamefont {J.~D.}\ \bibnamefont
  {Hunter}},\ }\href {http://link.aip.org/link/?CSX/9/90/1} {\bibfield
  {journal} {\bibinfo  {journal} {Comput. Sci. Eng.}\ }\textbf {\bibinfo
  {volume} {9}},\ \bibinfo {pages} {90} (\bibinfo {year} {2007})},\ \bibinfo
  {note} {\url{http://matplotlib.org}}\BibitemShut {NoStop}%
\bibitem [{\citenamefont {P\'erez}\ and\ \citenamefont
  {Granger}(2007)}]{ipython}%
  \BibitemOpen
  \bibfield  {author} {\bibinfo {author} {\bibfnamefont {F.}~\bibnamefont
  {P\'erez}}\ and\ \bibinfo {author} {\bibfnamefont {B.~E.}\ \bibnamefont
  {Granger}},\ }\href {http://link.aip.org/link/?CSX/9/21/1} {\bibfield
  {journal} {\bibinfo  {journal} {{C}omput. {S}ci. {E}ng.}\ }\textbf {\bibinfo
  {volume} {9}},\ \bibinfo {pages} {21} (\bibinfo {year} {2007})},\ \bibinfo
  {note} {\url{http://ipython.org}}\BibitemShut {NoStop}%
\end{thebibliography}%

\end{document}


\section{}

The measurement of the moments of the differential cross section $\mathrm{d}
\sigma / \mathrm{d} \cos \theta_t$ naturally contains correlations that are
important when performing fits to the measurement.  Table~\ref{eigen} provides
the eigenvalue decomposition of the covariance matrix, $C$.  In order to perform
fits to the measurement, one would compute the inverse covariance matrix,
$C^{-1}$, from the eigenvalue decomposition, and then minimize the negative
log-likelihood given by \begin{equation} \sum_{\ell, \ell'} (a_\ell -
	\hat{a}_\ell) (C^{-1})_{\ell\ell'} (a_{\ell'} - \hat{a}_{\ell'}),
\end{equation} where $a_\ell$ are the measured moments and $\hat{a}_\ell$ are
the moments predicted by the model being fit to the measurement.

\begin{table}
	\caption{The eigenvalues and eigenvectors of the covariance matrix for the
	parton-level measurement of $a_\ell$, the Legendre moments of the
    differential cross section $d \sigma / d \cos \theta_t$.  A single vertical
    column contains first an eigenvalue, then the error eigenvector
	corresponding to that eigenvalue.\label{eigen}}
\begin{ruledtabular}
\begin{tabular}{lrrrrrrrr}
Eigenvalue $\lambda$ &  0.452 &  0.254 &  0.159 &  0.0104 &  0.0276 &  0.0389 &  0.0975 &  0.0791 \\
\hline
$\ell = 1$           &  0.018 & -0.021 & -0.022 &  0.930 & -0.231 &  0.259 &  0.112 &  0.028 \\
$\ell = 2$           &  0.108 & -0.075 &  0.294 &  0.089 &  0.659 &  0.332 &  0.026 & -0.586 \\
$\ell = 3$           &  0.035 & -0.105 & -0.042 & -0.354 & -0.427 &  0.727 &  0.373 & -0.107 \\
$\ell = 4$           & -0.051 & -0.070 &  0.459 &  0.003 & -0.523 & -0.405 &  0.064 & -0.584 \\
$\ell = 5$           & -0.030 & -0.319 & -0.175 &  0.034 &  0.213 & -0.338 &  0.839 &  0.039 \\
$\ell = 6$           & -0.295 & -0.141 &  0.776 & -0.015 &  0.085 &  0.107 &  0.096 &  0.512 \\
$\ell = 7$           &  0.004 & -0.924 & -0.122 &  0.003 & -0.044 &  0.019 & -0.358 &  0.003 \\
$\ell = 8$           & -0.947 &  0.041 & -0.229 &  0.018 &  0.049 &  0.069 & -0.043 & -0.200 \\
\end{tabular}
\end{ruledtabular}
\end{table}